\documentclass[12pt]{iopart}
\usepackage{iopams, epsfig, amssymb, amsthm,psfrag, subfigure}
\usepackage{color}
\definecolor{Red}{rgb}{1,0,0}

\newcommand{\ua}{\uparrow}
\newcommand{\da}{\downarrow}

\newcommand{\beq}{\begin{equation}}
\newcommand{\eeq}{\end{equation}}
\newcommand{\ra}{\rangle}

\begin{document}

\title[Non-Abelian gauge fields]{Particles in non-Abelian gauge potentials \\
- Landau problem and insertion of non-Abelian flux}

\author{B Estienne, S M Haaker, K Schoutens}
\address{Institute for Theoretical Physics, University of Amsterdam,
Science Park 904, P.O.Box 94485, 1090 GL Amsterdam, The Netherlands}

\date{\today}

\begin{abstract}
We study charged spin-1/2 particles in two dimensions, subject to a perpendicular
non-Abelian magnetic field.  Specializing to a choice of vector potential that is
spatially constant but non-Abelian, we investigate the Landau level spectrum in
planar and spherical geometry, paying particular attention to the role of the
total angular momentum $\vec{J}=\vec{L}+\vec{S}$. After this we show that the
adiabatic insertion of non-Abelian flux in a spin-polarized quantum Hall state
leads to the formation of charged spin-textures, which in the simplest cases can be 
identified with quantum Hall Skyrmions.
\end{abstract}

\maketitle

\section{Introduction}

It has long been known that the motion of charged particles in a plane and in the presence
of a perpendicular magnetic field is highly special. In classical mechanics particles exhibit 
cyclotron motion, while quantum mechanics leads to the Landau level band structure
\cite{Landau}, with each 
level providing a macroscopic number of one-particle states that are strictly degenerate in energy. In this setting free electrons can form integer quantum Hall states, with intricate topological properties resulting in a quantization of the Hall conductance. Adding interactions leads to a plethora of remarkable many-body states, collectively known as fractional quantum Hall states. Such states are known to admit excitations with fractional charge and fractional statistics. The possibility that specific quantum Hall states have excitations with non-Abelian braid statistics has opened the exciting perspective of applications in the realm of topological quantum computation
\cite{Nayak}.

A setting that is physically very different but mathematically similar to that of 2D electrons in a perpendicular field is that of rapidly rotating cold atoms. In this analogy, the vorticity of a rotating liquid is akin to magnetic flux in the electron system. The limit of rapid rotation leads to a Landau level structure and one expects the formation of atomic quantum Hall states (see \cite{Cooper} for a review). The road towards experimental realization of atomic quantum Hall states is extremely challenging, but the lowest Landau level (LLL) has been reached  \cite{LLLexp} and there are indications of the formation of incompressible quantum liquids in small clusters \cite{Gemelke}.

The cold atom setting allows for yet another variation on the same theme: cold atomic gases subject to external time-dependent potentials that are such that they mimic the effects of magnetic fields or, equivalently, of rotation. Recent proposals for implementing such {\it artificial gauge fields} can be found in \cite{Dalibard,Jaksch,Mueller,Rosenkranz,Sorensen,Kolovsky,Juzeliunas1};
a successful experimental realization was reported in  \cite{Lin}.

Very interestingly, the latter setup is flexible enough to allow for a generalization of regular, Abelian, magnetic fields, corresponding to the $U(1)$ gauge symmetry of Maxwell's theory, to gauge fields pertaining to non-Abelian symmetries. The simplest version of this idea is the case of spin-1/2 particles subject to an external gauge potential for a non-Abelian $U(2)$ symmetry \cite{Ruseckas,Osterloh,Goldman,Goldman2}, see \cite{Juzeliunas2} for the case of spin-1 particles. 
Remarkably, very similar if not identical Hamiltonians arise in the original setting of a 2D electron gas in a perpendicular magnetic field, if Rashba \cite{Rashba} and Dresselhaus \cite{Dresselhaus} spin-orbit coupling terms are taken into account. This connection can be exploited to study the physics of the quantum spin Hall effect in the context of cold atoms \cite{Goldman3,Liu}.
 
With this, there is ample reason to study non-Abelian external gauge potentials with special emphasis on what is new as compared to the Abelian case. For particles confined to a lattice geometry, the non-Abelian case leads to an interesting generalization of the `Hofstadter butterfly' fractal 1-particle spectrum, dubbed `Hofstadter moth'  \cite{Osterloh}. The Landau level problem 
and the possibility of realizing (fractional) quantum Hall states with particles subject to non-Abelian gauge fields has already attracted a lot of attention \cite{Goldman,Goldman2,Burrello, Pachos}. 
 
In this paper we analyze non-interacting spin-1/2 particles in an external non-Abelian field and reflect on some of the fundamental differences with the Abelian setup. We analyze and solve the Landau level problem in spherical geometry, highlighting the fundamental role of the {\it total} angular momentum $\vec{J}=\vec{L}+\vec{S}$, which commutes with the Hamiltonian. In this setup, the non-Abelian field penetrating the sphere agrees with the asymptotic (large radius) limit of the  
non-Abelian magnetic monopoles first written by  't Hooft and Polyakov \cite{tHooftPol}.  One reason to focus on spherical geometry is that this geometry is known to be particularly useful for the purpose of a numerical study of many-body states arising upon adding interactions to the Landau level problem \cite{Haldane_sphere}.

We then proceed to the process where a non-Abelian flux is inserted in a background of otherwise Abelian flux. In our simplest case, $J_z=L_z+S_z$ remains a good quantum number during the flux insertion. This suggests a prominent role for transitions where particles flip their spin while at the same time changing their $L_z$ by (plus or minus) one unit, meaning that they jump to an adjacent Landau level orbital of the Abelian problem. This is indeed what happens: starting from a spin-polarized integer quantum Hall state, inserting non-Abelian flux at the origin leads to a state where, depending on their distance from the origin, particles have an amplitude for changing their spin and moving radially in or out. The resulting state is a spin-texture of unit electric and topological charge, which is easily identified with a quantum Hall Skyrmion. More general external fields lead to more intricate textures.

In the literature, thought experiments involving insertion of (Abelian) flux are often invoked as a probe of the characteristics of the quantum phase of a 
many body system. One spectacular example is the argument by Laughlin that insertion of a unit flux through a gapped medium with Hall conductance
$\sigma_H=\nu e^2/h$  leads to the nucleation of an excitation with fractional charge $e^*=\nu e$ \cite{Laughlin}. In the case of the quantum spin Hall state,
insertion of flux leads to spin-full excitations at the sample edges \cite{KaneMele}. One motivation for the present study has been the wish to extend these 
considerations to the case of non-Abelian flux. 

We briefly comment on possible experimental realization of our ideas. In the case of electrons, the integer quantum Hall reference state is readily available, 
and the non-Abelian flux can in principle, in the presence of spin-orbit coupling, be generated via a perpendicular electric field (see section 3 below). Integer quantum Hall states for cold atomic 
fermions are expected to arise in rapidly rotating systems \cite{Ho} or in laser-induced artificial gauge fields \cite{Ohberg}.
The paper \cite{Osterloh} has indicated how non-Abelian external gauge fields can be generated with the help of external lasers; if such laser pulses 
can spatially focussed, this will result in pulses of non-Abelian flux similar to the ones we propose here. Clearly these external fields will deviate from our 
expressions in their details, we expect however that the spin-textures that we predict here are relatively robust. Reading out spin-textures of effective 
spin-1/2 atomic states is possible; see for example \cite{Schweikhard}, where a Skyrmion lattice for effective spin-1/2 bosons was imaged.  

This paper is organized as follows. We start (section \ref{sec:nA}) with a general exposition on non-Abelian gauge fields. Next (section \ref{sec:LL}) we 
discuss the non-Abelian Landau problem, with particular emphasis on spherical geometry. Section \ref{sectNAflux} discusses the response of 
spin-full particles to the insertion of non-Abelian flux in an otherwise Abelian background. We end with some conclusions in section \ref{sec:conclusion}. The appendices A-D present further details and background material.

\section{On non-Abelian gauge fields}
\label{sec:nA}

Before we start to investigate the Landau problem, we start by a quick review on non-Abelian gauge fields, highlighting the differences with the Abelian case. We are interested in the 
quantum problem of a particle coupled to an external non-Abelian gauge field $\vec{A}$. In a cold atom experiment  these fields are controlled by external lasers. 
Therefore in the present article we are not concerned with the dynamics of  these gauge fields, which are treated as control parameters. The Hamiltonian for a non-relativistic particle of 
mass $m$ in an external magnetic field is 
 \begin{equation}
H = \frac{1}{2m}(\vec{p}-  \vec{A})^2 \label{generic_Hamiltonian} \ . 
\end{equation}
In the non-Abelian setup each component of the vector potential $ \vec{ A}$ is a matrix 
\begin{equation}
\vec{A} =  A_x \vec{u}_x +  A_y \vec{u}_y +  A_z \vec{u}_z = A_a \vec{u}_a \ .
\end{equation}
The corresponding magnetic field (or curvature) $\vec{B}$ is
\begin{equation}
\vec{B} = \vec{\nabla} \times \vec{A} -  \frac{i}{\hbar} \vec{A} \times \vec{A}  \label{B_def} \ ,
\end{equation}
or  equivalently in components:
\begin{equation}
{ B}_a  =  \epsilon_{abc} \left( \partial_b { A }_c - \frac{i}{\hbar} { A}_b { A}_c \right)   \label{potential} \ .
\end{equation}
The first term is the usual curl, while the second term $ \vec{A} \times \vec{A}$ vanishes identically for Abelian gauge fields. However it is non-zero in the generic non-Abelian case, 
as the components $A_a$ of the potential may not commute. In contrast to the Abelian case, even a uniform potential $\vec{A}$ can produce a non-zero magnetic field. Introducing the covariant derivative is $\vec{D} = \vec{\nabla} -\frac{i}{\hbar}\vec{A} = \frac{i}{\hbar}(\vec{p}-\vec{A})$, the magnetic field takes the following compact form:
\begin{equation}
\vec{B} = \vec{D} \times \vec{A}  \ . \label{B_def_bis}
\end{equation}
A gauge transformation is simply a unitary transformation $U$ and a change of $\vec{ A}$, such that $\vec{D}$ transforms covariantly:
\begin{equation}
\vec{A} \to U \vec{A} U^{\dagger} + i \hbar  U \vec{\nabla}  U^{\dagger} \ .
\end{equation}
Then the following quantities transform covariantly (note that the magnetic field is no longer gauge invariant):
\begin{equation}
\vec{D} \to U \vec{D} U^{\dagger}, \qquad \vec{B}  \to  U \vec{B} U^{\dagger} \ ,
\end{equation} 
and the Hamiltonian \eref{generic_Hamiltonian} is clearly a covariant quantity:
\begin{equation}
H = \frac{1}{2m}(\vec{p}-  \vec{A})^2 = -\frac{\hbar^2}{2m} \vec{D}^2 \ .
\end{equation}
In the Abelian case, two field configurations $\vec{A}$ yielding the same magnetic field are necessarily equivalent up to a gauge transformation (on a simply connected space). This is no longer the case for non-Abelian gauge groups. In particular it is known from \cite{Brown} that there are two gauge-inequivalent kinds of non-Abelian vector potentials which produce a uniform $\vec{B}$ field (although this statement is completely general, we illustrate this for $\vec{B} = 2\sigma_z \vec{u}_z$):
\begin{itemize}
\item a commuting field with linear potential $\vec{A} = \frac{1}{2}\vec{B}\times \vec{r} = \sigma_z (-y,x,0)$, for which only $\vec{\nabla} \times \vec{A}$ contributes to the field strength. This case is Abelian in nature, in the sense that all components of the potential vector commute with each other.
\item a uniform non-commuting potential, for instance $\vec{A} = (-\sigma_y,\sigma_x,0)$, such that   $\vec{B} =- \frac{i}{\hbar} \vec{A} \times \vec{A}$. This is only possible in a non-Abelian gauge group. 
\end{itemize}
Although these two kinds of potential give rise to the same magnetic field, they lead to completely different physical properties. While the first kind induces a magnetic length, Aharanov-Bohm effect and Landau level discrete spectrum, the second one has no Aharanov-Bohm effect, and the spectrum of a particle in such an external field configuration is continuous.

\section{The non-Abelian Landau problem: particles in perpendicular uniform non-Abelian external field}
\label{sec:LL}

In the present article we focus on the simplest case of non-Abelian gauge fields, when $A_a$ and $B_a$ are $2\times2$  hermitian matrices. The gauge group is then $U(2) = U(1) \times SU(2)$ and decomposes into:
\begin{itemize}
\item an Abelian $U(1)$ part, namely the fields proportional to the identity matrix $\mathbb{I}$,   
\item and a non-Abelian $SU(2)$ component, whose fields are linear combinations of the Pauli matrices $\sigma_a$.
\end{itemize}
The $U(2)$ case is a natural choice as it is the simplest case allowing non-Abelian gauge fields. However there is a deeper reason to focus on $U(2)$ gauge fields. Very similar physics can arise in a 2D electron gas when taking into account relativistic corrections in the Pauli-Schr\"{o}dinger equation, such as the Thomas term
\begin{equation}
H_{T}  =  -\frac{q \hbar}{4 m^2 c^2}  \vec{\sigma}\cdot \left( \vec{E} \times \vec{p} \right) \ . \label{Thomas}
\end{equation}
This spin-orbit coupling term plays a crucial role in spintronics \cite{Engel}, and it mimics the effect of a non-Abelian gauge potential:
\begin{equation}
\vec{A} \sim \vec{E} \times \vec{\sigma} \ .
\end{equation}
In this section we study the quantum problem of a non-relativistic particle confined to a two dimensional manifold in the background of a uniform perpendicular $U(2)$ magnetic field. We present the spectra for two different geometries: the plane and the sphere.  It turns out that this Hamiltonian can be mapped exactly to the one of a 2D electron in a perpendicular $U(1)$ magnetic field, when the Thomas term is present and an additional perpendicular $U(1)$ electric field $\vec{E}$ is applied. 

\subsection{On the plane}
In order to set this problem on the plane, we consider a perpendicular, uniform magnetic field $\vec{B} = B_z \vec{u}_z$, where $B_z$ is a $2\times2$  hermitian matrix. We can always choose a basis where the matrix $B_z$ is diagonal:
\begin{equation}
B_z = B \mathbb{I}  + \frac{2}{\hbar} \beta'^2 \sigma_z = B \left( \mathbb{I} + 2 \beta^2 \sigma_z \right) \ .
\end{equation}
We introduced the pure number $\beta=\frac{\beta'}{l_m B}$ involving the magnetic length $l_m = \sqrt{\frac{\hbar}{B}}$ (for $B >0$). From now on we work with $\hbar  =1$. This magnetic field is a $U(2)$ matrix, and is a superposition of a $U(1)$ field $B$ and a $SU(2)$ field $2 \beta^2 B \sigma_z$. 
In view of the previous discussion, there is an ambiguity in the notion of a non-Abelian uniform magnetic field, and one has to specify the non-Abelian part of the potential $\vec{A}$. The first kind of potential $\vec{A} = \frac{1}{2}\vec{B}\times \vec{r}$ boils down to an Abelian $U(1)\times U(1)$ gauge group, and the physics is simply that of two non-interacting species of particles coupled to different Abelian magnetic fields. The second kind however, a constant and 
non-commutative potential, is much more interesting and leads to new physics \cite{Goldman,Goldman2,Burrello,Pachos}: 
\begin{equation}
\vec{ A} = \frac{B}{2} \left(\begin{array}{c} -y\mathbb{I} \\ x\mathbb{I} \\ 0 \end{array}\right) + \beta'  \left( \begin{array}{c}  - a \sigma_y \\  a^{-1}\sigma_x \\ 0 \end{array}\right) \ . \label{constant_gauge}
\end{equation}
The Hamiltonian describing a particle confined to a plane in this non-Abelian background 
\begin{equation}
H = \frac{1}{2m}(\vec{p}- \vec{A})^2 \label{H_plane} \ ,
\end{equation}
enjoys the translation symmetry of the plane. Since the non-Abelian part of $\vec{ A}$ is uniform, the magnetic translation operators are insensitive to the non-Abelian part, and have the usual expressions:
\begin{equation}
T_x  =  \left(-i \partial_x - \frac{y}{2l_m^2} \right)\ ,  \qquad T_y  =  \left(-i \partial_y + \frac{x}{2l_m^2} \right)\ ,
\end{equation}
which implies immediately  the Abelian Aharanov-Bohm effect:
\begin{equation}
[\vec{a} \cdot \vec{ T}, \vec{b} \cdot \vec{ T}] = - i\frac{(\vec{a} \times \vec{b})\cdot \vec{u}_z }{l_m^2} \mathbb{I}  \label{AH}\ . 
\end{equation}
The r.h.s. is simply the flux of the Abelian part of the magnetic field $B\mathbb{I}$ through the parallelogram delimited by the vectors $\vec{a}$ and $\vec{b}$, and the (Abelian) magnetic length scale is $l_m$. Only the Abelian part of the magnetic field is quantized, and the number of states in a given Landau level will only depend on the Abelian field strength $B$.

It turns out that the problem of a particle in such a non-Abelian external field can be mapped exactly to the Hamiltonian of two-dimensional electron in the presence of both Rashba and Dresselhaus spin-orbit interaction \cite{Pachos}, and it has been first solved in this context by Zhang \cite{Zhang}. 

Having in mind to solve this problem on the sphere in the next section, we demand rotational symmetry around the $z$ axis, and we focus on the symmetric gauge: 
\begin{equation}
\vec{ A} = \frac{B}{2} \left(\begin{array}{c} -y\mathbb{I} \\ x\mathbb{I} \\ 0 \end{array}\right) + \beta'  \left( \begin{array}{c}  -\sigma_y \\  \sigma_x \\ 0 \end{array}\right) \ . \label{symmetric_gauge}
\end{equation}
This symmetric case correspond to the absence of Dresselhaus interaction in \cite{Zhang}, and the Hamiltonian \eref{H_plane} is much simpler to solve in this case. 
Moreover it can be mapped to a Thomas term \eref{Thomas} in the presence of a perpendicular uniform electric field $\vec{E} \propto \beta' \vec{u}_z$. This Hamiltonian can be expanded as:
\begin{equation}
 H = \omega_c \left(  a^{\dagger}a + \sqrt{2}\beta (a^{\dagger}\sigma_+ + a\sigma_-) + \frac{1}{2}(1+2\beta^2) \right) \ , \label{H_plane_JC}
\end{equation}
where $a,a^{\dagger}$ are the usual annihilation and creation operators appearing in the Landau problem (see \ref{Landau_plane}). Up to a change of spin basis $U=\sigma_x$ this is nothing but the celebrated Jaynes-Cummings Hamiltonian, and it is straightforward to obtain its spectrum:
\begin{eqnarray}
E_0 & = & \omega_c \left( \frac{1}{2} + \beta^2 \right) \ , \\
E_n^{\pm} &= &\omega_c \left(  n  \pm \sqrt{2\beta^2n + \frac{1}{4}} + \beta^2 \right) \ .
\end{eqnarray}

\subsection{On the sphere}

It can be rather instructive to solve such a problem on a sphere instead of the plane. The surface of the sphere being finite, the degeneracy of the Landau levels becomes finite too, which is very interesting for numerics. Moreover the translation invariance of the plane is promoted to the rotational symmetry of the sphere, and the spectrum decomposes into $\textrm{SU}(2)$ multiplets. In the Abelian case this geometry was first solved in \cite{WuYang}, and later used by Haldane \cite{Haldane_sphere} in the context of the quantum Hall effect. In order to fix our notations we recall these main results in  \ref{Landau_sphere}.

\subsubsection{Field configuration}

A uniform perpendicular magnetic field implies the presence of a magnetic monopole at the center  of the sphere.  In the Abelian case (see \ref{Landau_sphere}), the corresponding potential $\vec{A}_{Ab}$ must have a singularity (Dirac string) somewhere on the sphere, for instance at the south pole $\theta=\pi$: 
\begin{equation}
\vec{ A}_{Ab} =  \frac{N_{\Phi}}{2} \frac{1-\cos(\theta)}{r\sin(\theta)}\vec{u}_{\phi}  \ . \label{Abelian_potential}
\end{equation}
When Dirac's quantization condition $N_{\Phi} \in \mathbb{Z}$ is satisfied, this singularity has no physical consequence as it can be moved around through gauge transformations  \cite{Dirac_quantization}. This is the well known quantization of the magnetic flux piercing the sphere, which must be an integer number $N_{\Phi}$ of flux quanta: 
\begin{equation}
\int_S \vec{B}\cdot \textrm{d}\vec{S} = 2 \pi N_{\Phi}  \qquad i.e. \qquad B_r = \frac{N_{\Phi}}{2 r^2}  \ .
\end{equation}
To this $U(1)$ potential \eref{Abelian_potential} we add the following $SU(2)$ component:
\begin{equation}
\vec{ A}(\alpha) =  \vec{A}_{Ab}  + \alpha \frac{\vec{r}\times\vec{\sigma}}{r^2}  \ .\label{NA_potential}
\end{equation}
Once again this correspond to a Thomas term \eref{Thomas} with a radial uniform electric field $\vec{E} \propto \alpha \vec{u}_r$, and
this will turn out to be the correct extension of the symmetric gauge on the plane \eref{symmetric_gauge}. The corresponding magnetic field is
\begin{equation}
B_r = \frac{N_{\Phi}}{2 r^2} - 2 \alpha(1-\alpha) \frac{(\vec{r}\cdot\vec{\sigma})}{r^3} \ .
\end{equation}
It is quite remarkable that the strength $\alpha$ of the non-Abelian field need not be quantized, as the vector potential $ \frac{\vec{r}\times\vec{\sigma}}{r^2}$ has no singularity on the sphere: $\alpha$ can be any real number. As the radial $U(1)$ field is created by a magnetic monopole, it is not very surprising that the $SU(2)$ counterpart involves a non-Abelian monopole. Indeed, the potential $\vec{A}= \alpha \frac{\vec{r}\times\vec{\sigma}}{r^2}$ is nothing but the large distance asymptotic of a true non-Abelian monopole \cite{tHooftPol}, and in this context it is well known that there is no need for a Dirac string, nor is there a singularity anywhere on the sphere.

\subsubsection{Hamiltonian and spectrum}

\begin{figure}
\centering
\includegraphics[width= .7\textwidth]{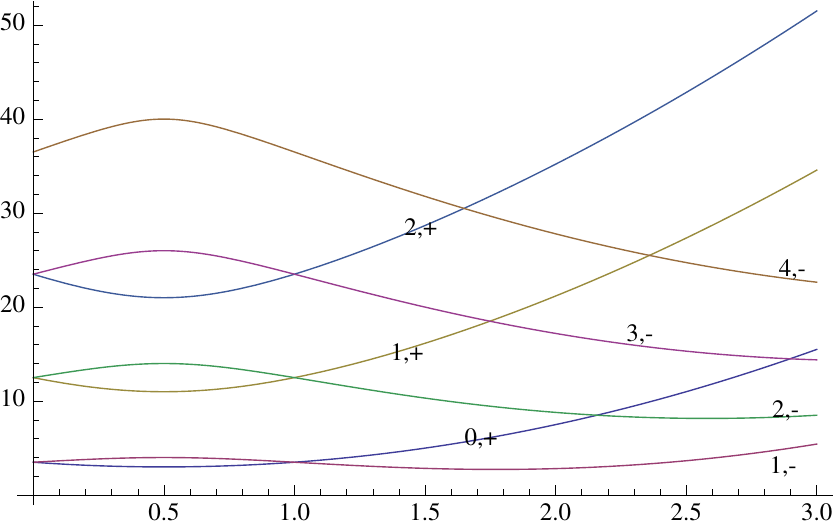}
\caption{Band structure for the non-Abelian Landau problem on the sphere as a function of the non-Abelian field strength $\alpha$. This is the case of $N_{\Phi} = 7$ Abelian flux quanta, and only the lowest part of the spectrum is shown.}
\label{sphere_spectrum_plot}
\end{figure}

The details about the derivation of the spectrum of a particle confined to a sphere of radius $r$ in this non-Abelian background can be found in \ref{sphere_spectrum}. The Hamiltonian is the following
\begin{equation}
H(\alpha) = \frac{1}{2m r^2} \left[ \vec{r} \times \left( \vec{p} - \vec{A}(\alpha) \right) \right]^2 \label{sphere_Hamiltonian}\ ,
\end{equation}
which is  a scalar under global rotations generated by $\vec{J} = \vec{L}+ \vec{S}$, where $\vec{L}$ is the angular momentum and $\vec{S}=\frac{1}{2}\vec{\sigma}$ is the spin. Its eigenstates form $SU(2)$ multiplets corresponding to the decomposition of the Hilbert space into irreducible representations of $\vec{J}$
\begin{equation}
\mathcal{H} =\left(j_0\right) \oplus 2  \left(j_1\right) \oplus 2  \left(j_2\right) \oplus \cdots \oplus 2\left(j_n\right)  \oplus \cdots   \label{Hilbert_space_U(2)} \ ,
\end{equation}
where $j_n = \frac{N_{\Phi}-1}{2} + n$. The corresponding eigenvalues are (see figure \ref{sphere_spectrum_plot}): 
\begin{eqnarray}
\hspace{-2.2cm}
E_0(\alpha) & = & \frac{1}{2m r^2} \left( \frac{N_{\Phi}}{2} -2\alpha(1-\alpha) \right) \ ,\\
\hspace{-2.2cm}
E^{\pm}_n(\alpha) & = & \frac{1}{2m r^2} \left( n(N_{\Phi}+n) -2\alpha(1-\alpha) \pm \sqrt{ \left(2\alpha - 1\right)^2 n\left( N_{\Phi}+n \right) + \left(\frac{N_{\Phi}}{2}\right)^2 } \right)  .
\end{eqnarray}
As can be seen in \fref{sphere_spectrum_plot}, multiple level crossings are observed. This also occurs in the planar case \cite{Burrello}, which can be recovered from the sphere in the limit of infinite radius.

\subsubsection{Recovering the plane}

The non-Abelian $\vec{A}$ field we considered on the sphere is indeed the correct extension of the planar symmetric gauge \eref{symmetric_gauge}. The planar problem is recovered by taking the sphere radius $r\to \infty$ while keeping constant the gauge field strength on the surface:
\begin{equation}
\frac{N_{\Phi}}{2r^2}\sim   B ,  \qquad \frac{\alpha}{r}  \sim \beta'=  \frac{\beta}{\sqrt{B}} \ .
\end{equation}
The vector potential and the magnetic field become in this limit:
\begin{equation}
\vec{A}  \to  \frac{B}{2} \left( \begin{array}{c} -y\mathbb{I} \\ x\mathbb{I} \\ 0 \end{array}\right) + \beta'  \left( \begin{array}{c}  -\sigma_y \\  \sigma_x \\ 0 \end{array}\right) 
\qquad  \qquad \vec{ B}  \to  B \left(  \mathbb{I} + 2\beta^2 \sigma_z \right) \vec{u}_z \ .
\end{equation}
It is straightforward to check that the eigenvalues of the Hamiltonian $E_{n}^{\pm}(\alpha)$ behave in the planar limit as:
\begin{eqnarray}
E_0(\alpha) & \to &  \omega_c \left( \frac{1}{2} + \beta^2 \right) \ ,\\
E^{\pm}_n(\alpha) & \to & \omega_c \left(  n  \pm \sqrt{2\beta^2n + \frac{1}{4}} + \beta^2 \right) \ .
\end{eqnarray}
reproducing the planar spectrum. Moreover one can expand the Hamiltonian on the sphere in terms of $\vec{L}$
\begin{equation}
{H} = \frac{1}{2mr^2}\left[ \vec{L}^2 -\left( \frac{N_{\Phi}}{2}\right)^2 + 2\alpha\left(  \vec{L} + \frac{N_{\Phi}}{2} \frac{\vec{r}}{r} \right) \cdot\vec{\sigma}  + 2\alpha^2\right] \ ,
\end{equation}
and using the Holstein-Primakoff representation:
\begin{eqnarray}
L_+ & = & b^{\dagger} \sqrt{N_{\Phi} +2a^{\dagger}a - b^{\dagger}b} \ , \\
L_- & = & \sqrt{N_{\Phi} + 2a^{\dagger}a - b^{\dagger}b} \, b  \ ,\\
L_z & = & b^{\dagger}b -\frac{N_{\Phi}}{2} - a^{\dagger}a \ ,
\end{eqnarray}
we recover the planar Hamiltonian:
\begin{equation}
{H} \to  \omega_c\left[ \left( a^{\dagger} a + \frac{1}{2}\right) +  \sqrt{2}\beta \left( a \sigma + a^{\dagger} \sigma_+ \right) + \beta^2 \right]  \ .
\end{equation}
This is not surprising in view of the mapping of this problem with a spin-1/2 electron under an effective non-Abelian potential coming from the Thomas term with a perpendicular electric field. Indeed, the infinite radius limit of a sphere in a radial $\vec{E}$ and $\vec{B}$ field is clearly the plane under perpendicular $\vec{E}$ and $\vec{B}$ fields.

In summary, we analyzed the non-Abelian Landau problem on the sphere and obtained the spectrum. The degeneracy of Abelian Landau levels is preserved, and the number of states per area remains $1/l_m^2$, as we expected from the absence of a non-Abelian Aharanov-Bohm effect on the plane. Indeed, the Abelian Aharanov-Bohm  \eref{AH} effect enforces a minimum surface of $l_m^2$ for the wavefunctions.

\section{Adiabatic insertion of non-Abelian flux}\label{sectNAflux}

For the non-Abelian gauge group $SU(2)$, there is no notion of a `quantum of non-Abelian flux'. 
To appreciate this fact, we recall that in an Abelian gauge theory the number of flux quanta is nothing but the winding number of the potential around the flux tube, and this integer cannot be continuously deformed. The homotopy group of $U(1)$ is the set of integers $\mathbb{Z}$. However, the situation is different for the gauge group $SU(2)$, which  has trivial homotopy 
group so that any flux insertion can be smoothly deformed away.  This explains that there is 
no non-Abelian analog of  the notion of flux quantum, in agreement with the absence of quantization of the non-Abelian field strength on the sphere.

Nonetheless, we wish to consider the insertion of non-Abelian flux in a quantum Hall (qH) fluid.
In the Abelian case the celebrated Laughlin argument shows that the insertion of a quantum of
Abelian flux leads to the accumulation of electric charge $\pm \nu e$, with $\nu$ the filling 
fraction of the qH liquid. Here we wish to start from an (integer) qH fluid and to insert, at given
location ${\cal O}$,  a non-Abelian field configuration $\delta \vec{A}$ centered at ${\cal O}$ and 
chosen such that (i) it generates no magnetic field away from ${\cal O}$ and (ii) it can be
removed by a gauge transformation. After this process we perform a gauge transformation on the evolved state, in such a way that it lives in the same Hilbert space as the state we started off with.

Before entering into details we briefly sketch the well known argument for the Abelian case, as
discussed by Laughlin \cite{Laughlin} and Halperin \cite{Halperin}. This leans heavily upon the existence of a gap separating the ground state of the qH liquid from its excited states. Take a qH droplet and adiabatically insert a flux quantum at the origin, $\delta A_\phi=\frac{-1}{r}$. This value of the flux is very special in the sense that it creates no Aharanov-Bohm effect and therefore it can be gauged away. This means that after this adiabatic insertion one can map the system back to its original field configuration. The net effect of this process is a charge transport away from the origin leaving behind a quasihole, as single particle states in the lowest Landau level  (see \ref{Landau_plane}) evolve according to
\begin{equation}
| m \rangle \to |m+1 \rangle \ .
\end{equation}
The usual description of adiabatic processes involves Berry phases \cite{Berry,Simon}, or more generally Berry matrices \cite{Wilczek} in the presence of ground state degeneracies. Before we specify our system in detail we first give a quick review on the derivation of the Berry matrix.

\subsection{Berry matrix}

Considering an adiabatic process in the presence of energy degeneracies the Berry phase should be generalized as was done in \cite{Wilczek}. We will give the details on the Berry matrix that are needed for our purposes. Suppose that the Hamiltonian is a smooth function of a parameter $\{\lambda(t)\}$ and at every point in parameter space has a degenerate ground state energy separated from higher levels by some finite gap. Starting out in a state $|\alpha(0)\rangle$ belonging to the lowest energy subspace of the total Hilbert space $\mathcal{H}_{E(0)}$, the adiabatic theorem tells us we end up in an eigenstate which is again an element of the subspace of ground states at time $t$, $|\alpha(t)\rangle$. During this process the eigenstates obey the Schr\"odinger equation $H(\lambda)|\alpha(\lambda)\rangle= E_\alpha(\lambda)|\alpha(\lambda)\rangle$. Since nothing forbids this state to pick up a phase or a unitary matrix the final state can be written as 
\beq
|\psi_\alpha(t)\rangle= e^{-\frac{i}{\hbar}\int_0^t E_0(t^\prime)dt^\prime}U_B(t)|\alpha(t)\rangle,
\eeq
where $U_B$ is the {\it Berry matrix} generalizing the Berry phase. It is a unitary mapping \mbox{$U_B(t): \mathcal{H}_{E_0(t)} \rightarrow \mathcal{H}_{E_0(t)}$} from the subspace of ground states to itself such that $U_B(0)=\mathbb{I}$. Such a map is known as a holonomy, cf. \cite{Simon}. The phase in front of the Berry matrix is the dynamical phase, depending on the evolution of the ground state energy, but it will not be of any importance for our consideration, so we will discard it. The Berry matrix can be written in terms of path ordered integrals
\begin{eqnarray}
U_B(t)&=\mathcal{P}\exp \left[i\int_0^t \mathcal{A}(t^\prime)dt^\prime \right]=\\
&=\mathbb{I}+\sum_{n=1}^\infty i^n\int_0^t dt_n \int_0^{t_n} dt_{n-1} \ldots \int_0^{t_2}dt_1 \mathcal{A}(t_1) \ldots \mathcal{A}(t_n),
\end{eqnarray}
where $\mathcal{A}_{\alpha,\beta}\equiv i\langle\alpha(t)|\frac{d}{dt}|\beta(t)\rangle$ is the Berry connection,  which behaves under unitary transformations as a gauge potential.

\subsection{Choosing the field configuration}

We consider a non-relativistic spin-1/2 particle confined to the plane in a perpendicular magnetic field, $B_z=B\mathbb{I}$. Writing the vector potential we use cylindrical coordinates and choose the symmetric gauge
\beq\label{symmgauge}
A_\phi= \frac{Br}{2}\mathbb{I}.
\eeq
The Hamiltonian of the system $H=\frac{1}{2m}(\vec{p}-q\vec{A})^2\mathbb{I}$ acts identically on both spin states, i.e. the Landau levels are doubly degenerate. 
We only consider the lowest Landau level (LLL) and write for the eigenstates $|m,\epsilon\rangle$ where $\epsilon\in\{\uparrow, \downarrow \}$. For more details on these eigenstates we refer to \ref{Landau_plane}. 
To derive the effect of inserting a field configuration $\delta\vec{A}(r)$ in this system, we will consider two particular field configurations labeled as $M=0$ and $M=-1$, respectively. 
In \ref{NAflux} we present a generic configuration, labeled by an integer $M$, of which these two are specific cases. There we also give a more detailed derivation, to avoid any cumbersome equations in the main body of this paper and we explain how the label $M$ can be interpreted. 
Mimicking the insertion of Abelian flux briefly mentioned at the start of this section, we will insert a gauge field in such a way that no additional magnetic field is created away from the origin. Furthermore, we choose a symmetric gauge and make the simplification $\partial_z(\delta \vec{A})=\vec{0}$. The field now looks like a pure gauge
\beq
\delta\vec{A}=iU(\lambda)\nabla U^\dagger(\lambda),
\eeq
for some unitary matrix $U(\lambda)$, which depends on a parameter $\lambda$ controlling the adiabatic process. The evolved Hamiltonian is now easily found to be 
\beq\label{hamilt}
H(\lambda)= U(\lambda)H(0)U^\dagger(\lambda).
\eeq
Since this is just a gauge transformation we automatically meet the requirement that there is a gap of $\hbar \omega_c$ separating the subspace of ground states from excited states at every point in parameter space. Also the eigenstates of \eref{hamilt} are easily found to be $U(\lambda)|m, \epsilon\ra$. This is all the information we need to construct the Berry matrix and find the evolved state. We separately present the results for the gauge configurations with $M=0$ and $M=-1$. 

\subsubsection{The case $M=0$.}

Our $M=0$ non-Abelian field configuration reads as follows
\begin{equation}\label{vecpot}
\begin{array}{ll}
\delta A_r(\lambda)=&\frac{-\lambda}{1+(\lambda r)^2}\sigma_\phi \\[2mm]
\delta A_\phi(\lambda) =& \frac{-\lambda^2r}{1+(\lambda r)^2}\sigma_z+\frac{\lambda}{1+(\lambda r)^2}\sigma_r
\end{array}
\end{equation}
where we introduced cylindrical coordinates and $\sigma_r = \vec{\sigma} \cdot \vec{u}_r$ and $\sigma_{\phi} = \vec{\sigma} \cdot \vec{u}_\phi$. The reason why we label this field by $M=0$ is stated in \ref{NAflux} and will become especially clear for the case $M=-1$. Note that for  $r \gg 1/\lambda$ this field configuration does not depend on $\lambda$ and behaves as
\beq\label{limitvecpot0}
\delta A_\phi \sim -\frac{1}{r}\sigma_z +\Or(1/r^2) \ .
\eeq
The structure of this field corresponds to shifting the orbital of a spin-$\ua$ (spin-$\da$) particle by $+1$ ($-1$), precisely what would happen by inserting an Abelian flux quantum, where the sign depends on the spin of the particle. From this point onwards, we will refer to such a field as a $\sigma_z$-flux quantum, to explicitly distinguish between it and an insertion of a purely Abelian flux. 

Our starting point is a fully polarized integer qH state, represented as a product state $|\psi(0)\ra=\bigotimes_{m=0}^{m_f}|m \ua \ra$, which has the the first $(m_f+1)$ LLL orbitals filled with spin-$\uparrow$ particles. We adiabatically insert the non-Abelian flux \eref{vecpot} by slowly sweeping 
$\lambda$ from $\lambda=0$ to its final value. Just like in the Abelian case we gauge this evolved state back to the initial situation. In this way the final state lives in the same Hilbert space as the initial one, $\mathcal{H}_{E(0)}$. The resulting final state is as follows
\beq\label{finalstate}
|\psi_0(\lambda) \rangle=\bigotimes_{m=0}^{m_f} \left(u_m(\lambda)|m \ua\ra-v_m(\lambda)|m+1 \da\ra\right).
\eeq
\psfrag{A}{\small $v_m(0.1)$}
\psfrag{B}{\small $u_m(0.1)$}
\psfrag{C}{\small $v_m(10)$}
\psfrag{D}{\small $u_m(10)$}
\begin{figure}
\begin{center}
\includegraphics[width= .7\textwidth]{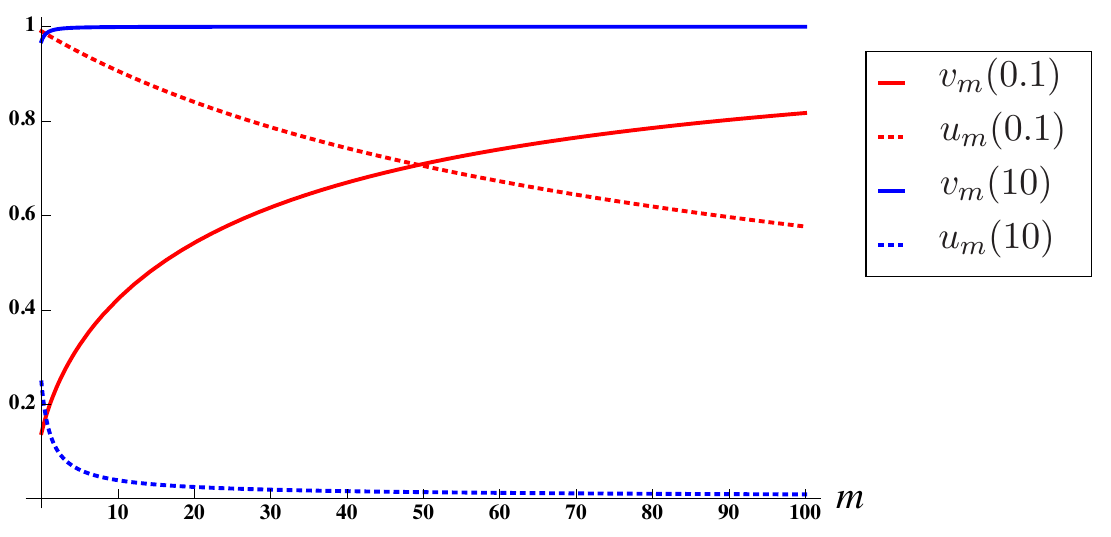}
\end{center}
\caption{Mixing coefficients $u_m(\lambda)$ and $v_m(\lambda)$, for two different values of $\lambda$. Around $m\sim 1/(2\lambda^2)$, $u_m(\lambda)$ and $v_m(\lambda)$ are equal to each other.}
\label{angles0}
\end{figure}
The mixing coefficients $\{u_m(\lambda),v_m(\lambda)\}$ depend on both the orbital $m$ and the adiabatic parameter $\lambda$. They are given in \eref{angleM} and are plotted in \fref{angles0} as a function of $m$, for two values of $\lambda$. Around orbital number $m\sim 1/(2\lambda^2)$, 
$u_m(\lambda)$ and $v_m(\lambda)$ cross, resulting in a vanishing of the $z$-component of the spin. 

The asymptotic behavior of the mixing coefficients as $m \to \infty$ can be read off from \eref{mlimitangle}. For large $m$, corresponding to a radius $r\gg1/\lambda$, the adiabatic process boils down to $| m \ua \ra \to | m+1 \da \ra$. This follows directly from the Berry matrix calculation, but it can be understood simply from the conservation of $J_z = L_z + \frac{1}{2} \sigma_z$. At large distance the flux we insert is essentially a $\sigma_z$-flux quantum \eref{limitvecpot0}, and it induces a charge transfer  $|m \ra \to |m+1\ra$, since all particles are spin-$\ua$, changing the angular momentum $L_z$ by one unit. Then the only way to accommodate the conservation of $J_z$ is through an accompanying 
spin flip $|\ua \ra \to |\da \ra$.

\begin{figure}
\centering
\includegraphics[width= .7\textwidth]{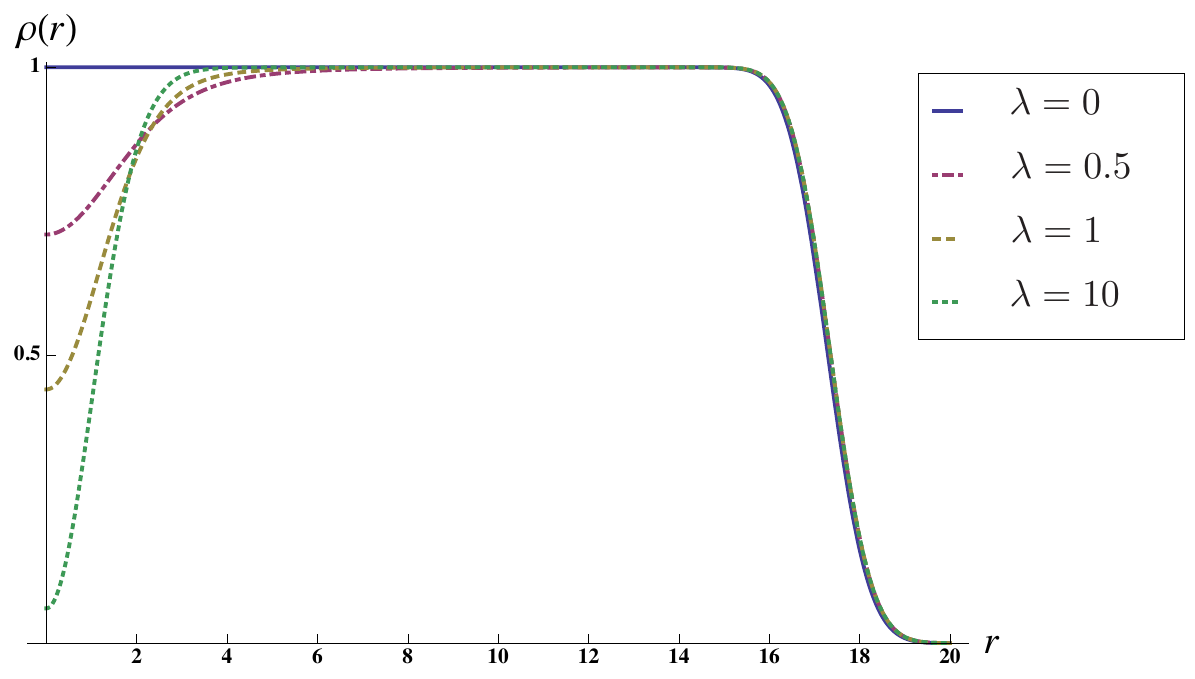}
\caption{Density profile for different values of $\lambda$ of a product state where the first 150 orbitals are filled with spin-$\uparrow$ particles. Before flux insertion ($\lambda=0$), there is a flat profile, but for finite values of $\lambda$, a quasihole of unit charge is created around the origin.}
\label{density0}
\end{figure}

We can analyze the effect of this adiabatic insertion on the product state, by looking at the density and spin profile of the final state \eref{finalstate}. The density is given by
\beq
\rho(r;\lambda)=\sum_{m=0}^{m_f}\frac{r^{2m} e^{-r^2/2}}{2^mm!2\pi}\left(u_m(\lambda)^2+v_m(\lambda)^2\frac{r^2}{2(m+1)}\right),
\eeq
and is shown in \fref{density0} for four different values of $\lambda$. The solid line shows the droplet before insertion of the non-Abelian field configuration, this is a flat profile. Once we insert flux, charge is depleted from the origin and deposited at the edge of the droplet. This is exactly one unit of charge. 

\psfrag{A}{\small $\lambda=0$}
\psfrag{B}{\small $\lambda=0.5$}
\psfrag{C}{\small $\lambda=1$}
\psfrag{D}{\small $\lambda=10$}

\begin{figure}
\centering
\includegraphics[width= .7\textwidth]{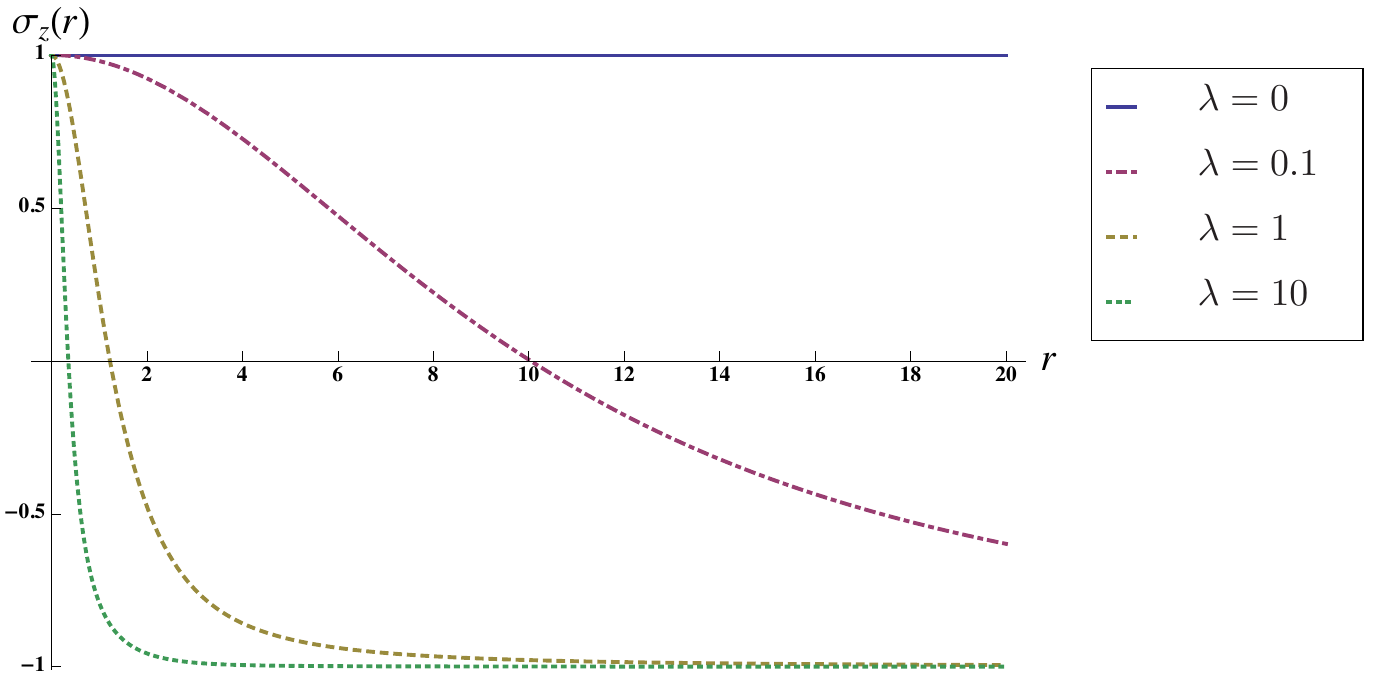}
\caption{
Expectation value of the the $z$-component of the spin field in a product state where the initial state has its first 150 orbitals filled with spin-$\uparrow$ particles. These profiles are after insertion of non-Abelian flux, for different values of $\lambda$. An insertion of finite $\lambda$ creates a non-trivial spin-texture. The radius at which the expectation value of $\sigma_z$ equals zero is around $r\sim1/\lambda$.}
\label{spinz0}
\end{figure}

\begin{figure}
\centering
\includegraphics{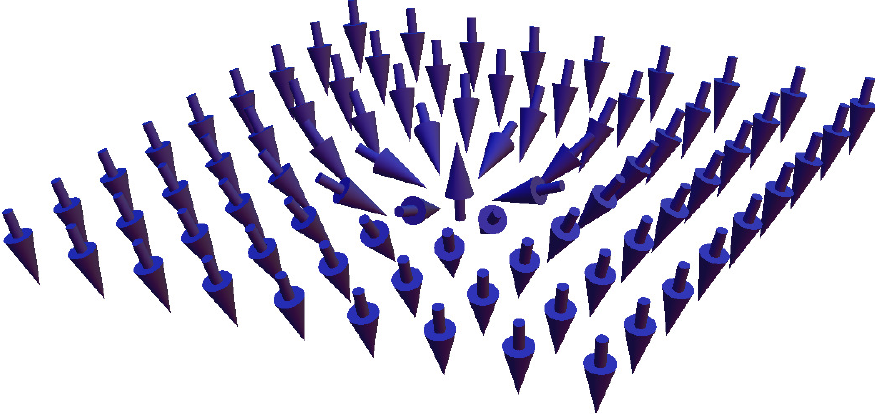}
\includegraphics[scale=0.6]{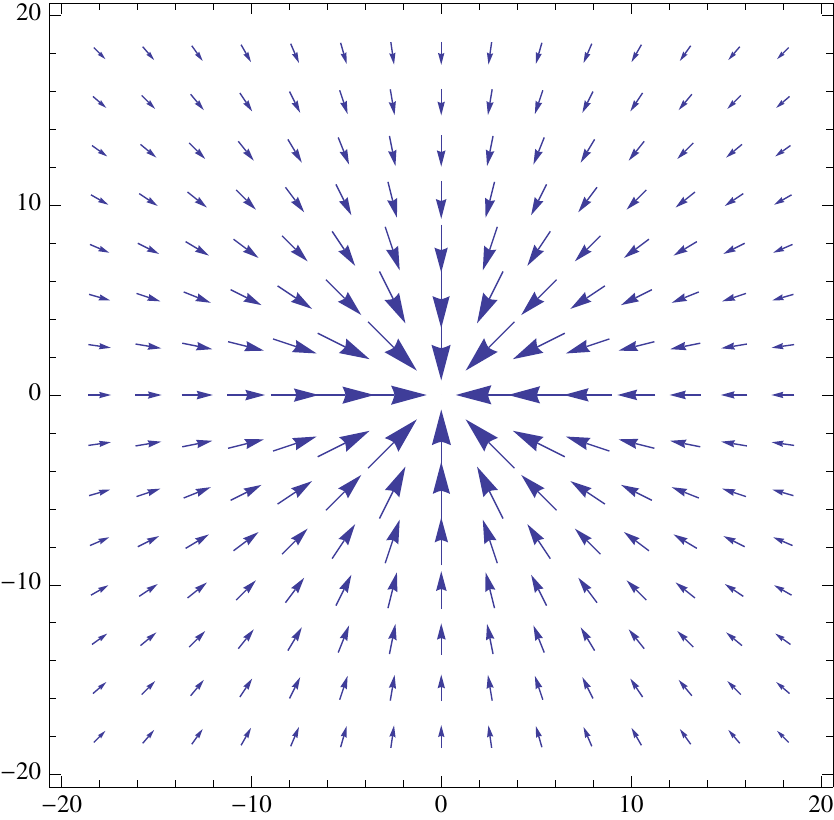}
\caption{Spin-texture obtained after inserting non-Abelian flux with $\lambda=1/3$. The three components of the spin field are shown in the left picture. In the origin the spin points up and at the edge it points down. The figure at the right depicts the $x$- and $y$-components of the spin field. These two figures clearly show that a Skyrmion with in-plane winding number 1 is created.}
\label{vectorplot0}
\end{figure}

The expectation value of spin in the $z$-direction of these configurations is depicted in figure \ref{spinz0}. The state before flux insertion is the blue solid line, which has trivial spin-texture. Upon increasing $\lambda$, particles move one orbital out while flipping their spin. This motion starts 
at the outer edge of the sample and propagates towards the center. When the final value of 
$\lambda$ has been reached,  the particles constitute a spin-texture of size $1/\lambda$
with spin-$\uparrow$ at the origin and spin-$\downarrow$ at the edge of the droplet.  Figure \ref{vectorplot0} shows the spin field after inserting a flux parameterized by $\lambda=1/3$.  The right picture displays the $(x,y)$-components of the spin field, showing that the spins have in plane winding number $1$.

The charged spin-texture created by the insertion of non-Abelian flux is recognized as a 
quantum Hall Skyrmion of unit electric charge, $q=e$, and unit topological charge, 
$Q_{\rm top}=1$. The topological charge, given by the Pontryagin index 
$Q_{\rm top}$, measures the winding of the spin vector around the system
(see, for example, \cite{Ezawa} chapter 7) and can easily be determined by looking at \fref{vectorplot0}. The left figure shows that the spin in the $z$-direction points up in the origin and down at the edge, so the in plane winding cannot be deformed into a trivial texture.

\psfrag{A}{\small $\lambda=0$}
\psfrag{B}{\small $\lambda=0.1$}
\psfrag{C}{\small $\lambda=1$}
\psfrag{D}{\small $\lambda=10$}

\subsubsection{The case $M=-1$.}

\begin{figure}
\centering
\includegraphics[width= .7\textwidth]{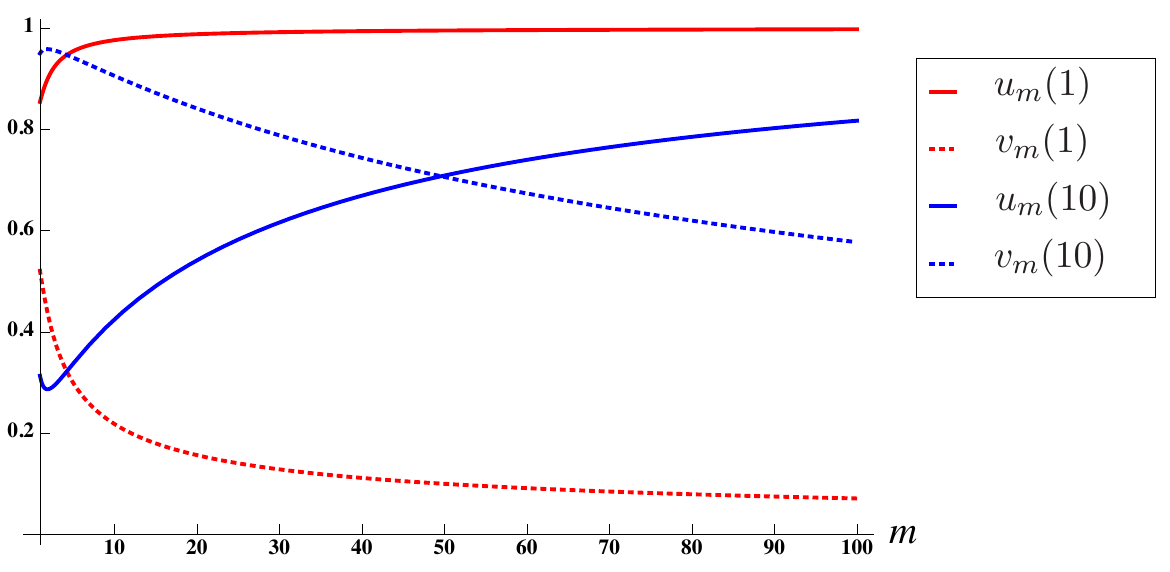}
\caption{Mixing coefficients of \eref{finalstate-1} for two different values of the adiabatic parameter $\lambda$. The point where $u_m(\lambda)=v_m(\lambda)$ is around $m\sim \lambda^2/2$. The asymptotes of the coefficients are exactly opposite from the $M=0$ case.}
\label{angles-1}
\end{figure}

Our second non-Abelian field configuration corresponds to the $M=-1$ case of the generic flux presented in \ref{NAflux}. We will insert it adiabatically into the initial setting \eref{symmgauge}. 
The gauge potential reads
\beq\label{field-1}
\begin{array}{lll}
\delta A_r(\lambda) &= &\frac{\lambda}{\lambda^2+r^2}\sigma_\phi\nonumber\\[2mm]
\delta A_\phi(\lambda) &=& -\frac{r}{\lambda^2+r^2}\sigma_z-\frac{\lambda}{\lambda^2+r^2}\sigma_r.
\end{array}
\eeq
There is a subtlety which did not arise in the previously discussed configuration and which will shed light on why we label the different fields by an integer $M$. This time $\delta A_\phi(0)=\frac{-1}{r}\sigma_z\neq0$, which is the insertion of a $\sigma_z$-flux quantum, resulting in a shift of orbital number depending on the spin of the particle
\beq
|m \ua\ra \rightarrow |m+1 \ua\ra \ , \qquad |m \da\ra \rightarrow |m-1 \da\ra.
\eeq
The adiabatic process consists of two parts now. We start by adiabatically inserting a $\sigma_z-$-flux quantum, leading to the configuration \eref{field-1} at $\lambda=0$. After that we slowly sweep $\lambda$ so as to reach its final value. Note that at every point of the adiabatic process we are able to find the eigenstates of the evolved Hamiltonian. 
Again starting from a product state of spin-$\uparrow$ particles and gauging back to the original Hamiltonian after the adiabatic process we get a final state
\beq\label{finalstate-1}
|\psi_{-1}(\lambda) \rangle=\bigotimes_{m=0}^{m_f} \left(u_{m+1}(\lambda)|m+1 \ua\ra-v_{m+1}(\lambda)|m \da\ra\right),
\eeq
where the coefficients can be found in \eref{angleM} and are plotted in \fref{angles-1}. This time the scale at which the spins are flipped is set by $r\sim\lambda$. 
\psfrag{A}{\small $u_m(1)$}
\psfrag{B}{\small $v_m(1)$}
\psfrag{C}{\small $u_m(10)$}
\psfrag{D}{\small $v_m(10)$}

\begin{figure}
\centering
\includegraphics[width= .7\textwidth]{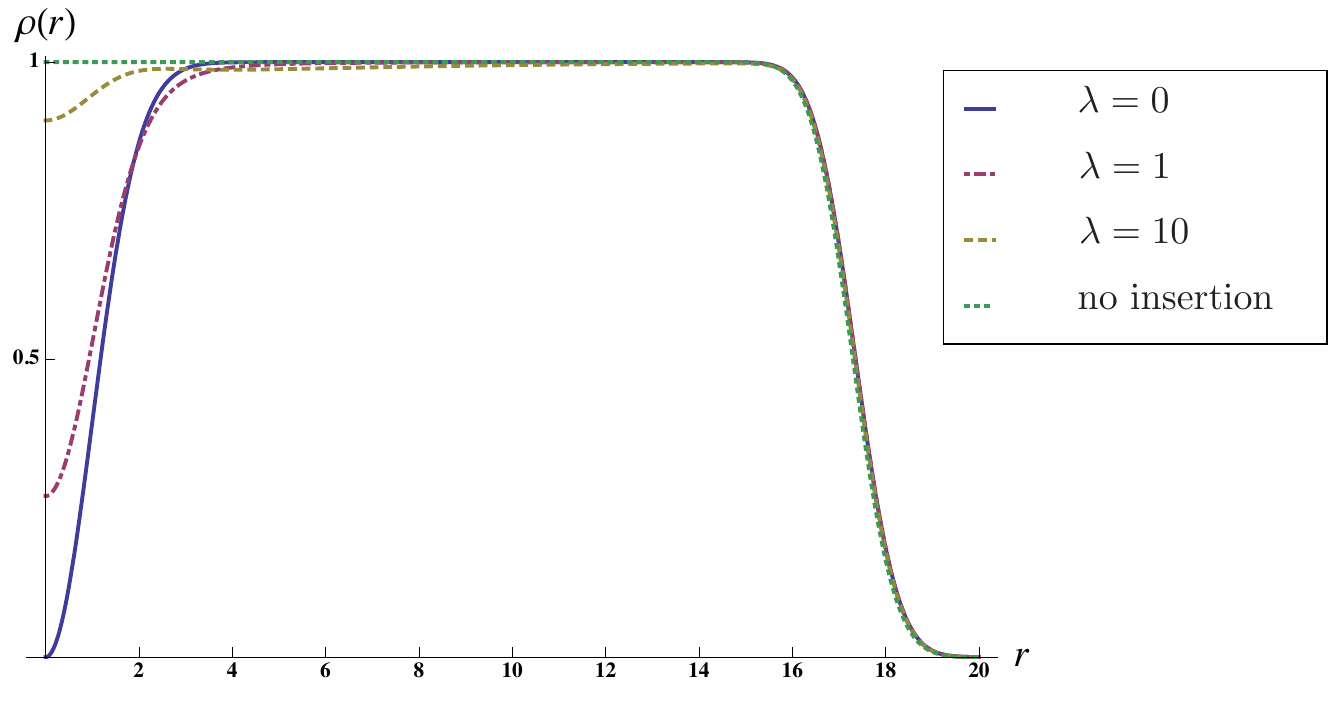}
\caption{Density of a product state before flux insertion and after for different values of $\lambda$. The flat profile depicts the initial product state. For $\lambda=0$ a $\sigma_z$-flux quantum is inserted creating a quasihole at the origin. At finite $\lambda$, there is still a density depletion  around the origin, but it is less sharp.}
\label{density-1}
\end{figure}

\begin{figure}
\centering
\includegraphics[width= .7\textwidth]{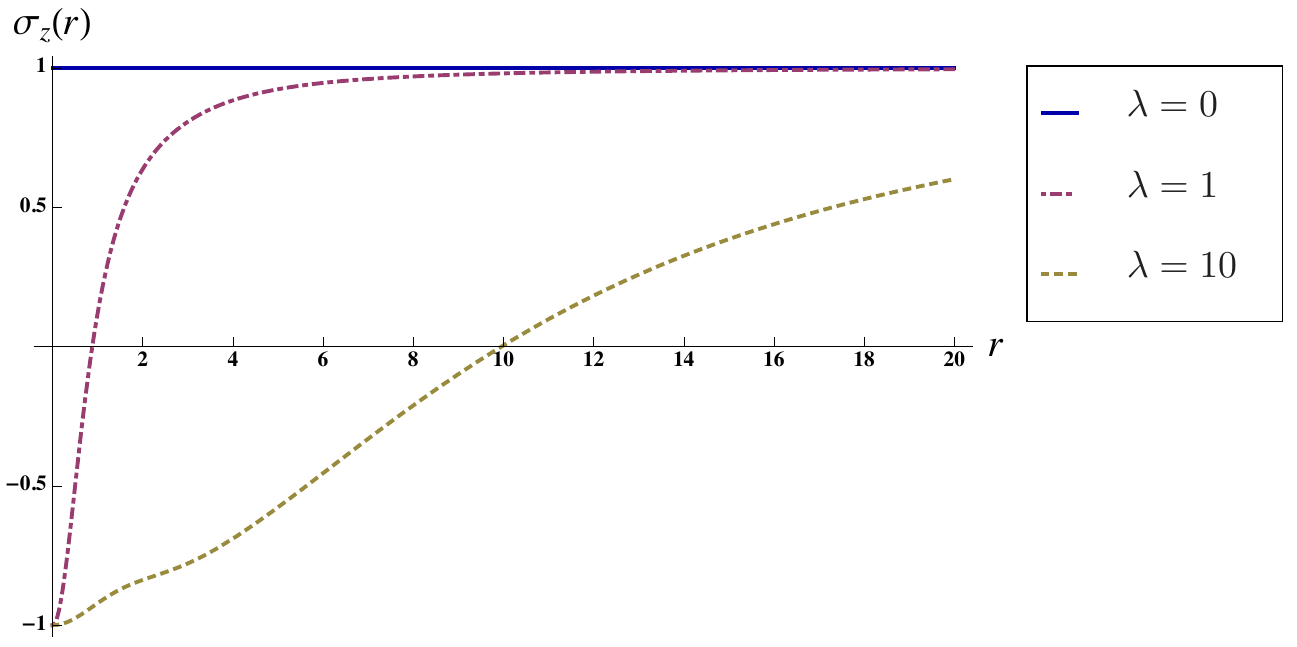}
\caption{Expectation value of the $z$-component of the spin field, in a product state \eref{finalstate-1} for different values of $\lambda$. The radius at which the state is unpolarized is around $r\sim \lambda$.}
\label{spinz-1}
\end{figure}

\begin{figure}
\centering
\includegraphics{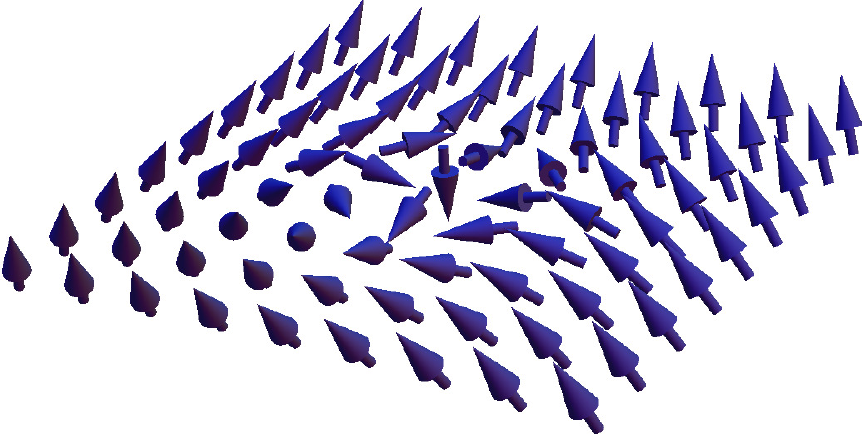}
\includegraphics[scale=0.6]{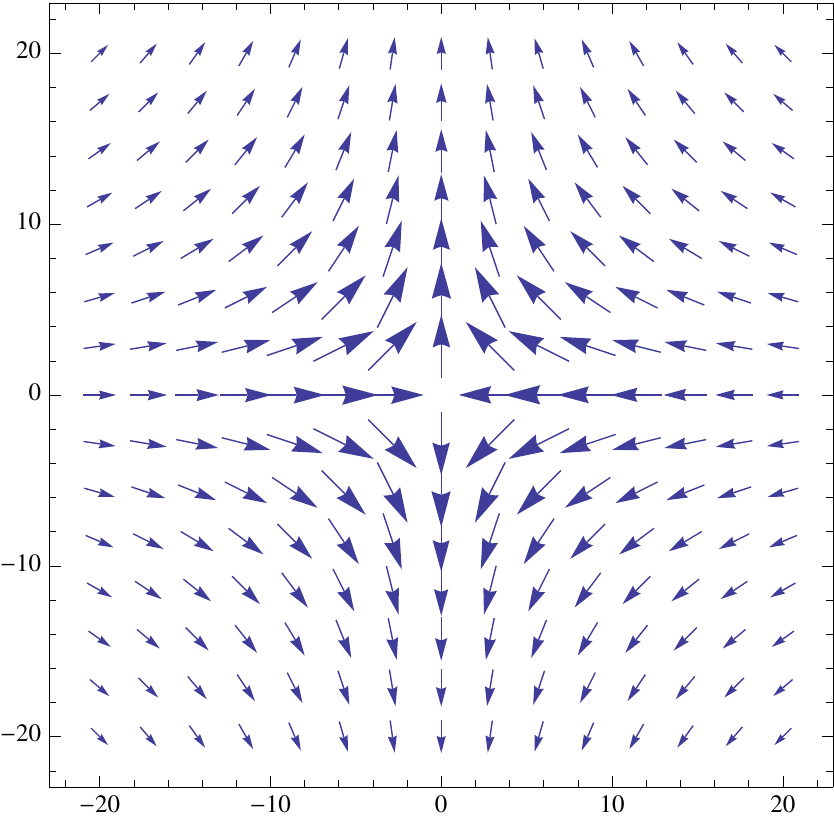}
\caption{The left figure shows the spin field for a final state labeled by $\lambda=5$. In the origin the spin is pointing down, at the edge it points up. The right figure shows $x$- and $y$-components of the spin field, from which we see that this flux insertion created a Skyrmion with in-plane winding number $-1$.}
\label{vectorplot-1}
\end{figure}

The density of \eref{finalstate-1} is shown in \fref{density-1}, before flux insertion and for the values $\lambda=0, 1,10$. Again charge is depleted from the origin, but this time the depth of the hole is largest for $\lambda=0$, meaning after the insertion of a $\sigma_z$-flux quantum at $\lambda=0$. Upon increasing $\lambda$, particles move inward while flipping their spin. This motion starts at the origin and moves out towards the edge of the sample. The resulting spin-texture is depicted in
\fref{spinz-1} and \fref{vectorplot-1}. The electric charge ($q=e$) and topological charge
$Q_{\rm top}=1$ agree with those of the Skyrmion found for $M=0$.

\psfrag{D}{\small no insertion}
\psfrag{A}{\small $\lambda=0$}
\psfrag{B}{\small $\lambda=1$}
\psfrag{C}{\small $\lambda=10$}

\section{Conclusion}
\label{sec:conclusion}

In this paper we analyzed and solved the non-Abelian Landau problem on the sphere and analyzed the charge and spin dynamics induced by the insertion of non-Abelian flux in an otherwise Abelian background. 

We remark that, in the usual (Abelian) quantum Hall setting, quantum Hall Skyrmions arise due to a balance between the effects of the Zeeman energy, which favors single overturned spins, and the Coulomb interaction, which favors configurations with small spin-gradients \cite{Sondhi}.  It is quite remarkable that our procedure of driving the non-interacting polarized electron gas with non-Abelian external flux leads to the very same Skyrmion configurations.

Repeating the non-Abelian flux insertion in a background of a $\nu=2$ integer quantum Hall 
state, with both the spin-$\uparrow$ and spin-$\downarrow$ lowest Landau levels completely filled, has a very different effect. In this case, the bulk state cannot accommodate any spin-flips and the effects 
of the flux-insertion are limited to the edges. Inserting non-Abelian flux through the central 
hole in a Corbino disc leads to neutral $S_z=\pm 1$ excitations at both inner and outer edge. 
This situation is in many ways reminiscent of a thought experiment where a minimal amount of Abelian flux inserted into a 2D spin quantum Hall (SQH) topological phase acts as a spin pump, resulting in neutral $S_z=\pm1/2$ excitations at the edges \cite{KaneMele}.

The details of the charge and spin dynamics associated to the insertion of non-Abelian flux depend on the specific form of our gauge potentials and on the way these depend on the sweep-parameter $\lambda$. One expect that many features, in particular the topological quantum numbers characterizing the resulting spin-textures, will be robust against changes in the detailed shape of the external gauge potentials.   

\appendix

\section{Landau levels on the plane}\label{Landau_plane}

For the purpose of being self-contained, and also in order to fix notations, we recall the main results about the Landau problem on the plane. We consider a particle of charge $q$ and mass $m$ confined to a plane, under an external perpendicular magnetic field $\vec{B}= B \vec{u}_z$ (with $qB>0$). The standard choice for the vector potential $\vec{A}$ is:
\begin{equation}
\vec{A} = \frac{B}{2}\left( \begin{array}{c} -y \\ x \\0 \end{array} \right) = \frac{Br}{2} \vec{u}_{\phi} .
\end{equation}
This is called the symmetric  gauge because it behaves as a vector under rotations around $\vec{u}_z$.
The only scale of the classical problem is the cyclotron frequency $\omega_c$:
\begin{equation}
\omega_c = \frac{qB}{m} .
\end{equation}
The quantum mechanical problem has an additional scale, the magnetic length $l_m$:
\begin{equation}
l_m=\sqrt{\frac{\hbar}{qB}} .
\end{equation}
The Hamiltonian reads in the symmetric gauge:
\begin{equation}
H = \frac{1}{2}\omega_c \left( \left(-il_m\partial_x + \frac{y}{2l_m} \right)^2 + \left(-il_m\partial_y - \frac{x}{2l_m} \right)^2 \right) .
\end{equation}
It is very convenient to go to complex coordinates (rescaled by the magnetic length), and to introduce two commuting families of creation and annihilation operators:
\begin{eqnarray}
a = \sqrt{2}  \left( \bar{\partial} + \frac{z}{4} \right) & \qquad & a^{\dagger} = \sqrt{2} \left( -\partial + \frac{\bar{z}}{4} \right) \\
b = \sqrt{2}\left( \partial + \frac{\bar{z}}{4} \right) & \qquad & b^{\dagger} = \sqrt{2} \left( -\bar{\partial} + \frac{z}{4} \right) .
\end{eqnarray}
where $\partial = \frac{\partial}{\partial z}$  and  $\bar{\partial}= \frac{\partial}{\partial \bar{z}}$. 
In these notations the Hamiltonian and angular momentum have a very simple expression:
\begin{equation}
H =  \omega_c \left( a^{\dagger} a + \frac{1}{2} \right) \ ,  \qquad L_z = b^{\dagger}b - a^{\dagger}a \ ,
\end{equation}
from which the spectrum $E_n = \omega_c (n+1/2)$ follows immediately. $b$ and $b^{\dagger}$ are also the generators of magnetic translation. Since they commute with the Hamiltonian, all these eigenvalues are infinitely degenerate. The subspace of energy $E_n = \omega_c (n+1/2)$ is called the n$^{th}$ Landau level (LL). 

Denoting by $n$ and $m$ the eigenvalues of $a^{\dagger}a$ and $b^{\dagger}b$ respectively, the Hilbert space is spanned by the states $|n,m\rangle$ for $n,m \geq 0$. The additional quantum number $m$ is related the value of the angular momentum $L_z | n,m \rangle = (m-n) |n,m\rangle $.

\begin{figure}
\centering
\includegraphics[width= .5\textwidth]{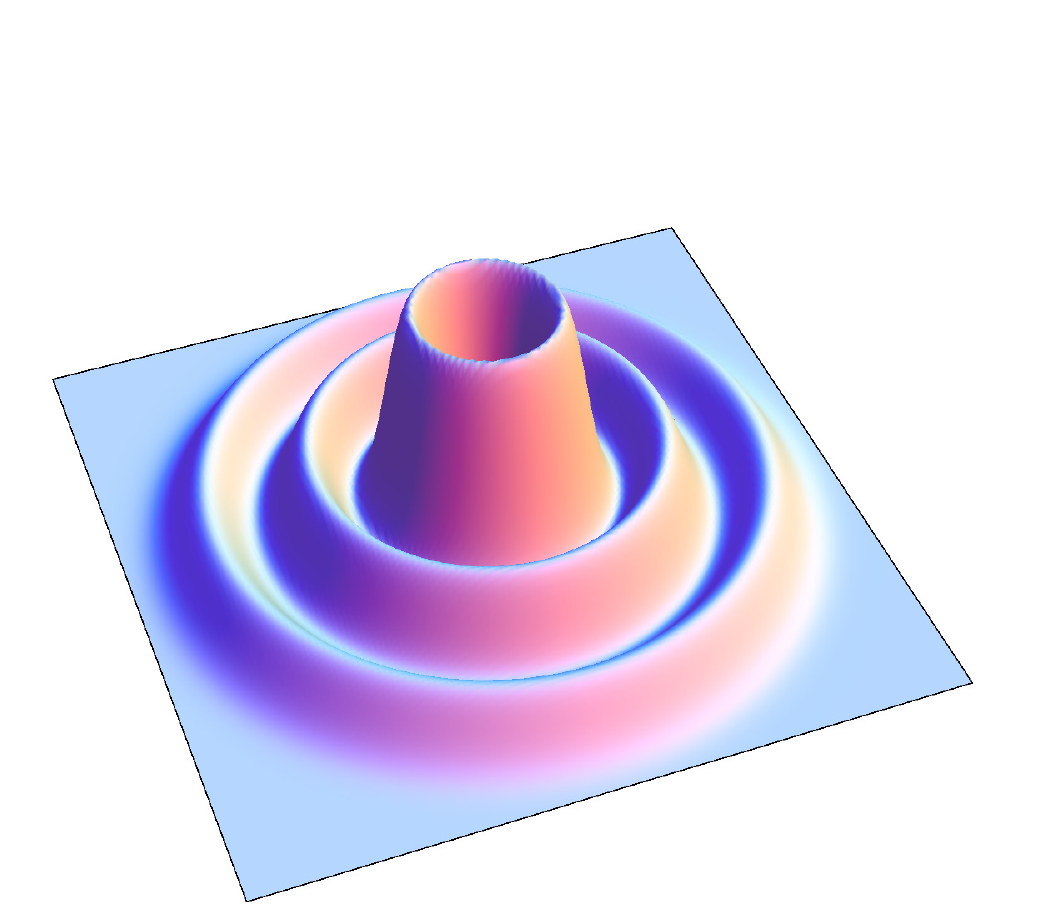}
\caption{Density profile of few orbitals in the lowest Landau level, namely $m=2,12$ and $30$ LLL orbitals. The cylindrical symmetry is clearly visible.}
\label{LLL_orbitals}
\end{figure}

The explicit form of their wave functions is known and involves a special class of functions called Hermite polynomials. In this appendix we focus on the lowest Landau level $n=0$:
it is obtained by acting with $b^{\dagger}$ on the state $|0,0\rangle$:
\begin{equation}
|0,m\rangle = \frac{\left(b^{\dagger}\right)^m}{\sqrt{m!}}|0,0\rangle 
\ \rightarrow\ \langle z | 0,n\rangle =   \frac{1}{\sqrt{2\pi}}\frac{z^m}{ \sqrt{2^m m!}} \exp(-z\bar{z}/4) \label{LLL_plane} \ .
\end{equation}
The orbital $|m\rangle$ has angular momentum $m$, and the support of the wave function \eref{LLL_plane} is a ring located  at distance $ \sqrt{2m} $ (in magnetic length scale) from the origin, as can be seen from \fref{LLL_orbitals}.

\section{Landau levels on the sphere}\label{Landau_sphere}

In this section of the Appendix we set $q=\hbar=1$. 

\subsection{Field configuration: magnetic monopole}

On the sphere, a uniform perpendicular magnetic field $\vec{B} = \frac{N_{\phi}}{2r^2} \vec{u}_r$ implies the presence of a magnetic monopole at the center  of the sphere, and  the  potential $\vec{A}$ must have a singularity (Dirac string) somewhere on the sphere. The gauge where the singularity lies at the south pole, i.e.
\begin{equation}
\vec{A} = \frac{N_{\Phi}}{2} \frac{1-\cos(\theta)}{r\sin(\theta)}\vec{u}_{\phi} \label{south}
\end{equation}
and the gauge where the singularity lies at the north pole:
\begin{equation}
\vec{A} =  -\frac{N_{\Phi}}{2} \frac{1+\cos(\theta)}{r\sin(\theta)}\vec{u}_{\phi} \label{north}
\end{equation}
are related by the following unitary transformation $U = e^{i N_{\phi} \phi}$. This operator only makes sense on the sphere when $N_{\Phi}$ is an integer. This is Dirac's quantization condition \cite{Dirac_quantization}.

\subsection{Hamiltonian and spectrum}

The Hamiltonian of a particle confined to the sphere of radius $r$ in the background of such a magnetic monopole is:
\begin{equation}
H = \frac{\vec{\Lambda}^2}{2mr^2}  \qquad \textrm{with} \qquad \vec{\Lambda} = \vec{r} \wedge \left( \vec{p} -  \vec{A} \right) .
\end{equation}
The operators $\Lambda_a$ have the following (gauge invariant) commutation relations
\begin{equation}
[\Lambda_a,\Lambda_b] = i \epsilon_{a b c} \left( \Lambda_c +  (\vec{r}\cdot\vec{B}) r_c\right) 
\end{equation}
and the generators of (magnetic) rotations have the form
\begin{equation}
\vec{L} = \vec{\Lambda} - (\vec{r}\cdot\vec{B}) \vec{r} = \vec{L} -  \frac{N_{\Phi}}{2}\vec{u}_r \label{L} .
\end{equation}
They generate a $\textrm{SU}(2)$ symmetry:
\begin{equation}
[L_a,L_b] = i \epsilon_{abc}L_c
\end{equation}
and the Hamiltonian boils down to the Casimir $L^2$. Indeed the relation:
\begin{equation}
\Lambda^2 = L^2 - \left(\frac{N_{\Phi}}{2}\right)^2 
\end{equation}
\begin{itemize}
\item ensures that all $L_a$ commute with the Hamiltonian.
\item gives the spectrum of the Hamiltonian: $\frac{1}{2mr^2} \left( l(l+1) - \left(\frac{N_{\Phi}}{2}\right)^2 \right)$
\end{itemize}
The last statement simply comes from the $\textrm{SU}(2)$ algebra obeyed by the $L_a$'s, which forces the eigenvalues of  $L^2$ to be of the form $l(l+1)$ where $l \in \frac{1}{2}\mathbb{N}$.  However not all these values of $l$ are possible. Using the explicit expression of $\vec{L}$, Wu and Yang \cite{WuYang} obtained the following decomposition of the Hilbert space into irreducible representations of the $SU(2)$ algebra generated by $\vec{L}$
 \begin{equation}
\mathcal{H} =\left(\frac{N_{\Phi}}{2}\right) \oplus   \left(\frac{N_{\Phi}}{2}+1\right) \oplus \cdots \oplus \left(\frac{N_{\Phi}}{2}+n\right)  \oplus \cdots   \label{Abelian_Hilbert}
\end{equation}
and the (Abelian) spectrum on the sphere finally reads:
\begin{equation}
E_n = \frac{1}{2mr^2}\left( n(N_{\phi}+n+1) + \frac{N_{\Phi}}{2}\right) \qquad n\geq 0 \ .
\end{equation}

\section{More details about the non-Abelian field on the sphere}\label{sphere_spectrum}

In this section we derive the spectrum of the Hamiltonian
\begin{equation}
H(\alpha) = \frac{1}{2m r^2} \left[ \vec{r} \times \left( \vec{p} - \vec{A}(\alpha) \right) \right]^2,
\end{equation}
describing a particle confined to a sphere of radius $r$ in the non-Abelian background potential 
\begin{equation}
\vec{ A}(\alpha) =  \vec{A}_{Ab}  + \alpha \frac{\vec{r}\times\vec{\sigma}}{r^2},
\end{equation}
where $\vec{A}_{Ab}$ stands for the $U(1)$ potential \eref{south}. 
Note that there is a gauge transformation mapping $\alpha \to 1-\alpha$ implemented by the unitary transformation $U=\frac{\vec{r}\cdot \vec{\sigma}}{r}=\sigma_r$.

\subsection{Rotational symmetry and decomposition of the Hilbert space}

There are two sets of $\textrm{SU}(2)$ generators in this problem:
\begin{itemize}
\item the usual (Abelian) action on the coordinates implemented by\\ $\vec{L} =  \vec{r} \times \left(\vec{p}- \vec{A}_{Ab}\right) - \frac{N_{\Phi}}{2}\frac{\vec{r}}r$ defined in \eref{L},
\item the rotations in spin space generated by $\vec{S} = \frac{1}{2}\vec{\sigma}$.
\end{itemize}
The Hamiltonian we are considering is not invariant under $\vec{L}$ and $\vec{S}$ separately. However, it is a scalar under global rotations generated by $\vec{J} = \vec{L}+ \vec{S}$, as can be seen from the expansion in terms of $\vec{J}$:
\begin{equation}
\hspace{-2cm}
 H(\alpha)  = \frac{1}{2m r^2} \left[ J^2 + \frac{1}{4} - 2\alpha(1 -\alpha)  + (2\alpha-1)\left(\vec{ J}\cdot\vec{\sigma} - \frac{1}{2}+  \frac{N_{\Phi}}{2} U\right) + \frac{N_{\Phi}}{2}U  \right].
\end{equation}
Therefore this Hamiltonian is block diagonal with respect to the decomposition of the Hilbert space into irreducible representations of $\vec{J}$. This decomposition follows directly from the Abelian one \eref{Abelian_Hilbert}:
\begin{equation}
\hspace{-1cm}
\mathcal{H} =\left(\frac{N_{\Phi}-1}{2}\right) \oplus 2  \left(\frac{N_{\Phi}+1}{2} \right) \oplus 2\left(\frac{N_{\Phi}+3}{2}\right) \oplus 2\left(\frac{N_{\Phi}+5}{2}\right)  \oplus \cdots   
\end{equation}

\subsection{Spectrum}

Working in the subspace $J^2=j(j+1)$, we simply need to diagonalize the term $X =  (2\alpha-1)\left(\vec{ J}\cdot\vec{\sigma} - \frac{1}{2}+  \frac{N_{\Phi}}{2} U\right) + \frac{N_{\Phi}}{2}U $.  We first derive the following two relations
\begin{eqnarray}
 \left\{ U,\left(\vec{J}\cdot\vec{\sigma} - \frac{1}{2} + \frac{N_{\Phi}}{2} U \right) \right\} & = & 0 \label{relation_anticommutation} \\
\left( \vec{J}\cdot \vec{\sigma} - \frac{1}{2}\right)^2 & = & \vec{J}^2 + \frac{1}{4} .
\end{eqnarray}
The first one is a consequence of the gauge equivalence $U H(\alpha) U = H(1-\alpha)$, and the second one can be checked using the explicit form of $\vec{J}$. From this we deduce that $X^2$ is a constant:
\begin{equation}
X^2  =   \left(2\alpha -1\right)^2 \left(\vec{J}^2 + \frac{1}{4}\right) + \alpha(1-\alpha)N_{\Phi}^2 \label{relation_square}
\end{equation}
and we get the following spectrum for $X$:
\begin{equation}
\lambda^{\pm}_{(j)} = \pm \sqrt{ \left(2\alpha - 1\right)^2 \left(\left(j+\frac{1}{2}\right)^2 - \left(\frac{N_{\Phi}}{2}\right)^2 \right) + \left(\frac{N_{\Phi}}{2}\right)^2 } .
\label{eigenvalues_U(2)}
\end{equation}
As can be seen in \eref{Hilbert_space_U(2)}, for $j \geq \frac{N_{\Phi}+1}{2}$ there are two representations of spin $(j)$, and from \eref{relation_anticommutation} both eigenvalues $\lambda^{\pm}_{(j)}$ belong to the spectrum. However there is a unique representation of spin $j = \frac{N_{\Phi}-1}{2}$. In this irrep. $U=1$ and the corresponding eigenvalue is the positive one: $\frac{N_{\Phi}}{2}$. Rewriting $j= n + \frac{N_{\Phi}-1}{2}$, we get the following spectrum for the Hamiltonian:
\begin{eqnarray}
\hspace{-2.5cm}
E_0(\alpha) & = & \frac{1}{2m r^2} \left( \frac{N_{\Phi}}{2} -2\alpha(1-\alpha) \right) \\
\hspace{-2.5cm}
E^{\pm}_n(\alpha) & = & \frac{1}{2m r^2} \left( n(N_{\Phi}+n) -2\alpha(1-\alpha) \pm \sqrt{ \left(2\alpha - 1\right)^2 n\left( N_{\Phi}+n \right) + \left(N_{\Phi}/2\right)^2 } \right) .
\end{eqnarray}

\section{Generic non-Abelian field configuration}\label{NAflux}

In this appendix, we give a detailed derivation of the final state obtained after an adiabatic insertion of non-Abelian flux. This is done for the generic case of which two specific examples are discussed in \sref{sectNAflux}. The field configuration we insert is the following
\beq\label{genericvecpot}
\begin{array}{lll}
\delta A_r(\lambda)&=& -\left(M+\frac{1}{2}\right)\frac{2\lambda r^{2M}}{1+\lambda^2r^{2+4M}}\sigma_\phi
\\[2mm]
\delta A_\phi(\lambda)&=& \left(M+\frac{1}{2}\right)\frac{1-\lambda^2r^{2+4M}}{1+\lambda^2r^{2+4M}}\frac{1}{r}\sigma_z+\left(M+\frac{1}{2}\right)\frac{2\lambda r^{2M}}{1+\lambda^2r^{2+4M}}\sigma_r-\frac{1}{2r}\sigma_z,
\end{array}
\eeq
where $M$ can be interpreted as the number of $\sigma_z$-flux quanta inserted which will be explained below \eref{genericeigenstates}.
Inserting such a field boils down to performing a gauge transformation on the system
\beq
U_M(\lambda)=\frac{1}{\sqrt{1+\lambda^2r^{2+4M}}}\left(\begin{array}{cc}
1&-\lambda\bar{z}r^{2M}\\
\lambda zr^{2M}&1
\end{array}\right)\exp(iM\phi\sigma_z).
\eeq
So at every point of the adiabatic process we know the LLL eigenstates of the evolved Hamiltonian, they are given by
\beq\label{genericeigenstates}
|\alpha(\lambda)\ra=U_M(\lambda)|m, \epsilon\ra.
\eeq
Before we proceed with calculating the Berry matrix an important subtlety needs to be considered. We wish to insert this field configuration into a background \eref{symmgauge}. But at $\lambda=0$ and for $M\neq0$ \eref{genericvecpot} is given by $\delta A_\phi(0)=(M/r)\sigma_z\neq0$, 
which means we have to start by adiabatically inserting $M$ $\sigma_z$-flux quanta, resulting into a shift of the orbitals depending on the spin of the particle
\beq
|m\ua\ra \rightarrow |m-M \ua\ra \ , \qquad |m\da\ra \rightarrow |m+M \da\ra.
\eeq
After the insertion of these $\sigma_z$-flux quanta, we slowly sweep $\lambda$ from zero to some final value resulting into \eref{genericvecpot}. Now we can use the eigenstates \eref{genericeigenstates} to compute the Berry connection
\beq
\mathcal{A}_{\alpha,\beta}\equiv i\langle\alpha(\lambda)|\frac{d}{dt}|\beta(\lambda)\rangle=i\langle\alpha(0)|U_M^\dagger\dot U_M|\beta(0)\rangle,
\eeq
where 
\beq
iU_M^\dagger \dot U_M=\frac{i\dot \lambda}{1+\lambda^2r^{2+4M}}\left(\begin{array}{cc}
0&-\bar{z}^{2M+1}\\
z^{2M+1}&0
\end{array}\right).
\eeq
The Berry connection only has non-zero elements between states of the form  \mbox{$\{ U_M(\lambda)|m \ua\ra,U_M(\lambda)|m+2M+1 \da\ra\}$}. Written in this basis, for every $m$ the Berry matrix is a \mbox{$2 \times 2$ matrix}
\beq
U_{B}^m=\cos(\theta_m^{(M)}(\lambda))\mathbb{I}+i\sin(\theta_m^{(M)}(\lambda))\sigma_y,
\eeq
where the angle is given by
\beq\label{angleM}
\theta_m^{(M)}(\lambda)=\int_0^\infty dr \arctan(\lambda r^{1+2M})\frac{r^{2+2M+2m}e^{-r^2/2}}{2^{m+M} \sqrt{2m!(m+2M+1)!}}.
\eeq
This angle has interesting asymptotes in two different limits
\begin{eqnarray}
\lim_{m\rightarrow \infty}\theta_m^{(M)}=\arctan(\lambda (2m)^{M+1/2})\label{mlimitangle} , \\
\lim_{\lambda \rightarrow \infty}\theta_m^{(M)}=\frac{\pi}{2} \frac{\Gamma(m+M+3/2)}{\sqrt{m!(m+2M+1)!}} \ .\label{lambdalimitangle}
\end{eqnarray}
After the adiabatic insertion of flux we gauge the system back to the initial one. This cycle has the following effect on a single particle state $|m \ua\ra$
\beq\label{state}
\hspace{-1cm}
 U_M^\dagger(\lambda)U_B^m(\lambda)U_M(\lambda) |m \ua\rangle=u_m^{(M)}(\lambda)|m \ua\ra-v_m^{(M)}(\lambda)|m+2M+1 \da\ra,
\eeq
where the mixing coefficients are expressed in terms of \eref{angleM}
\beq
u_{m}^{(M)}(\lambda)  \equiv  \cos(\theta_{m}^{(M)}(\lambda)) \ , \qquad
v_{m}^{(M)}(\lambda)  \equiv \sin(\theta_{m}^{(M)}(\lambda)) \ .
\eeq
That the equality in \eref{state} holds can be seen by inserting unity $U_M(\lambda)|m^\prime, \epsilon \rangle\langle m^\prime, \epsilon|U_M^\dagger(\lambda)$ between $U_M^\dagger$ and $U_B^m$. After deducing the effect of the two stages of the adiabatic process, we can combine them to find the final state. Before we give the final state, one last remark needs to be made. Since we want to stay in the LLL, we have to put the state on a Corbino disc, meaning that we fill the orbitals of the initial product state with spin-$\ua$ particles starting from some initial orbital $m_i$ up to a final orbital $m_f$. The two specific adiabatic flux insertions given in \sref{sectNAflux} are actually the only two situations for which the Corbino disc is not a necessary geometry for staying in the LLL. 
Starting from a product state on a Corbino disc where the orbitals are filled with spin-$\uparrow$ particles, the final state after first adiabatically inserting $M$ $\sigma_z$-flux quanta, then cranking
up the value of $\lambda$ in \eref{genericvecpot}, and finally gauging back to the initial configuration, is given by
\beq
|\psi_M(\lambda) \rangle=\bigotimes_{m=m_i}^{m_f} \left(u_{m-M}^{(M)}(\lambda)|m-M \ua\ra-v_{m-M}^{(M)}(\lambda)|m+1+M \da\ra\right).
\eeq

\section*{Acknowledgements}
We thank Sander Bais and Sasha Zozulya for inspiring discussions, and Victor Gurarie for pointing out the reference \cite{Brown}.
This work was supported in part by the foundation FOM of the Netherlands.

\end{document}